\documentstyle[11pt,newpasp,twoside,epsf]{article}
\def\edcomment#1{\iffalse\marginpar{\raggedright\sl#1\/}\else\relax\fi}
\reversemarginpar

\begin{document}
\title{RAVE: The RAdial Velocity Experiment}

\hspace{0.4cm}\vspace*{1\baselineskip}Matthias Steinmetz\footnote{for the RAVE Science Working Group}\\
\affil{Astrophysikalisches Institut Potsdam, An der Sternwarte 16, 14482 Potsdam, Germany}

\begin{abstract}

RAVE\footnote{http://www.aip.de/RAVE} (RAdial Velocity Experiment) 
is an ambitious program to conduct an all-sky 
survey (complete to V = 16) to measure the radial velocities, metallicities and 
abundance ratios of 50 million stars using the 1.2-m UK Schmidt Telescope of the
Anglo-Australian Observatory (AAO), together with a northern counterpart, 
over the period 2006 - 2010. The survey will represent a giant leap forward in our 
understanding of our own Milky Way galaxy, providing a vast stellar kinematic database 
three orders of magnitude larger than any other survey proposed for this coming decade. 
RAVE will offer the first truly representative inventory of stellar radial velocities 
for all major components of the Galaxy. 

The survey is made possible by recent technical innovations in
multi-fiber spectroscopy; specifically the development of the
'Echidna' concept at the AAO for positioning fibers using
piezo-electric ball/spines.  A 1m-class Schmidt telescope equipped
with an Echidna fiber-optic positioner and suitable spectrograph would
be able to obtain spectra for over $20\;000$ stars per clear night.  

Although the main survey cannot begin until 2006, a key component of
the RAVE survey is a pilot program of $10^5$ stars which may be
carried out using the existing 6dF facility in unscheduled bright time
over the period 2003--2005.  

\end{abstract}

\section{Introduction}

In the first decade of the 21st century, it is being increasingly
recognized that many of the clues to the fundamental problem of galaxy
formation in the early Universe lie locked up in the motions and
chemical composition of stars in our Milky Way galaxy (for a review,
see e.g. Freeman \& Bland-Hawthorn 2002).  Consequently, significant
effort has been placed into planning the next generation of
large-scale astrometric surveys like GAIA. Stellar spectroscopy plays
a crucial role in these studies, not only providing radial velocities
as a key component of the 6-dimensional phase space of stellar
positions and velocities, but also providing much-needed information
on the chemical composition of individual stars. Taken together,
information on space motion and composition can be used to unravel the
formation process of the Galaxy.

However, the GAIA mission, which will provide astrometry, radial
velocities and chemical abundance for up to 1 billion stars, is
unlikely to be completed before the end of the next decade. Among the
existing surveys, the HIPPARCOS and the Tycho-2 catalog have
compiled proper motions for 118\,000 and for 2.5 million stars,
respectively, but radial velocities have been completed for only a few
ten thousands. The new release of the HST
guide star catalog, GSC2.3, will include proper motions,
and the astrometric satellite DIVA plans to compile proper motions for
up to 40 million stars, but no radial velocities are available for
these targets. To our knowledge no systematic survey is planned so far
that includes radial velocities and that is capable of filling this
gap in size and time between existing surveys and the GAIA mission.

With the successful demonstration of ultra-wide-field (40 deg$^2$)
multi-object spectroscopy (MOS) on the UK Schmidt telescope of the
AAO, a major opportunity beckons to generate the first large-scale
all-sky spectroscopic survey of Galactic stars with a radial velocity
precision better than 2\,km\,s$^{-1}$ (see Fig.~1).

In this paper we map out the case for RAVE (RAdial Velocity
Experiment), an ambitious plan to measure radial velocities and
chemical compositions for up to 50 million stars by 2010 using novel
instrumentation techniques on the UK Schmidt telescope and on a northern
counterpart.

\section{The RAVE survey}

RAVE is a large international collaboration involving a still growing
list of astrophysicists from Australia, Canada, France, Germany,
Italy, Japan, the Netherlands, the UK and the USA. The RAVE survey is
split into two components: a pilot survey and a main survey.

\begin{figure}
\plottwo{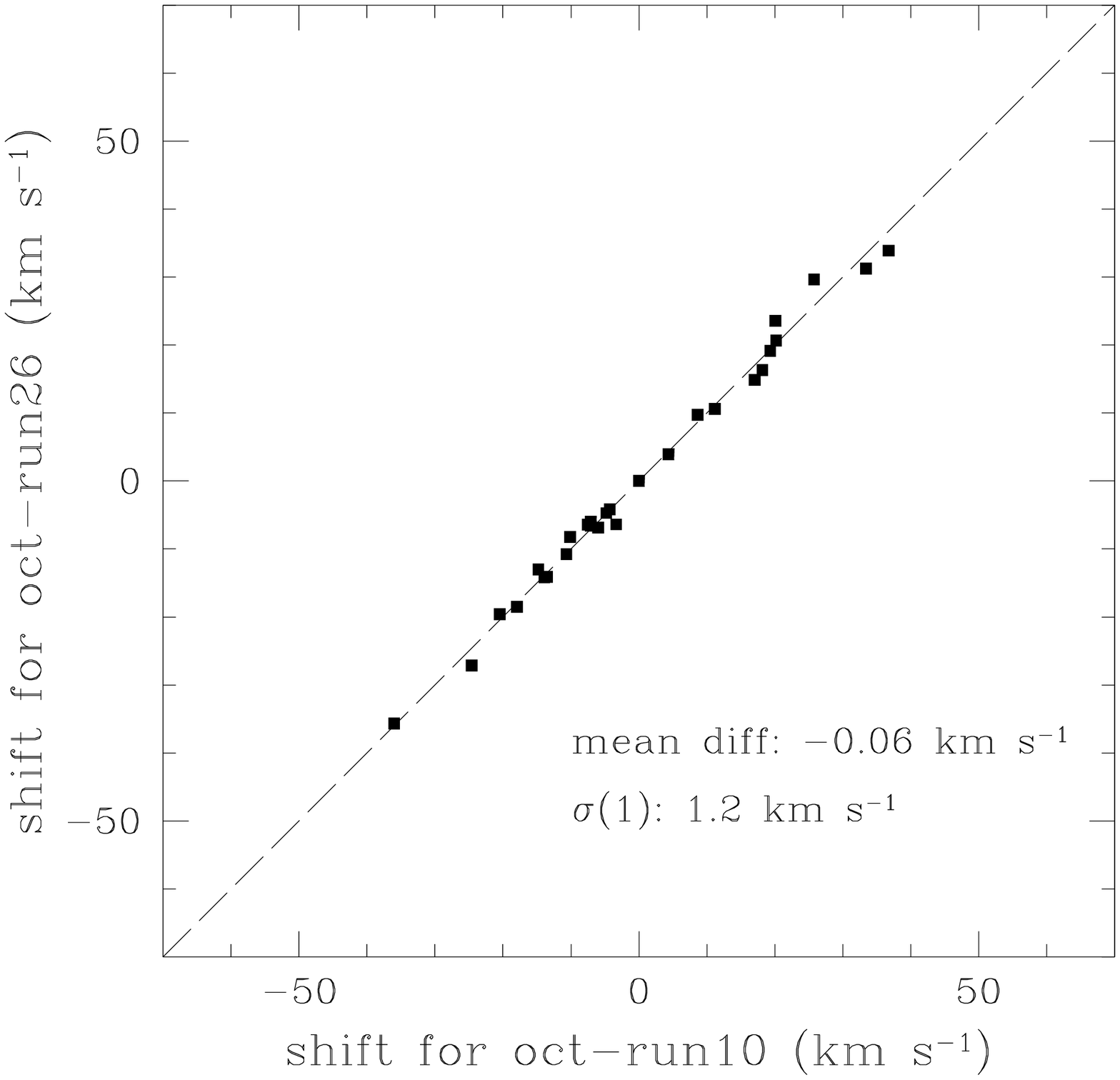}{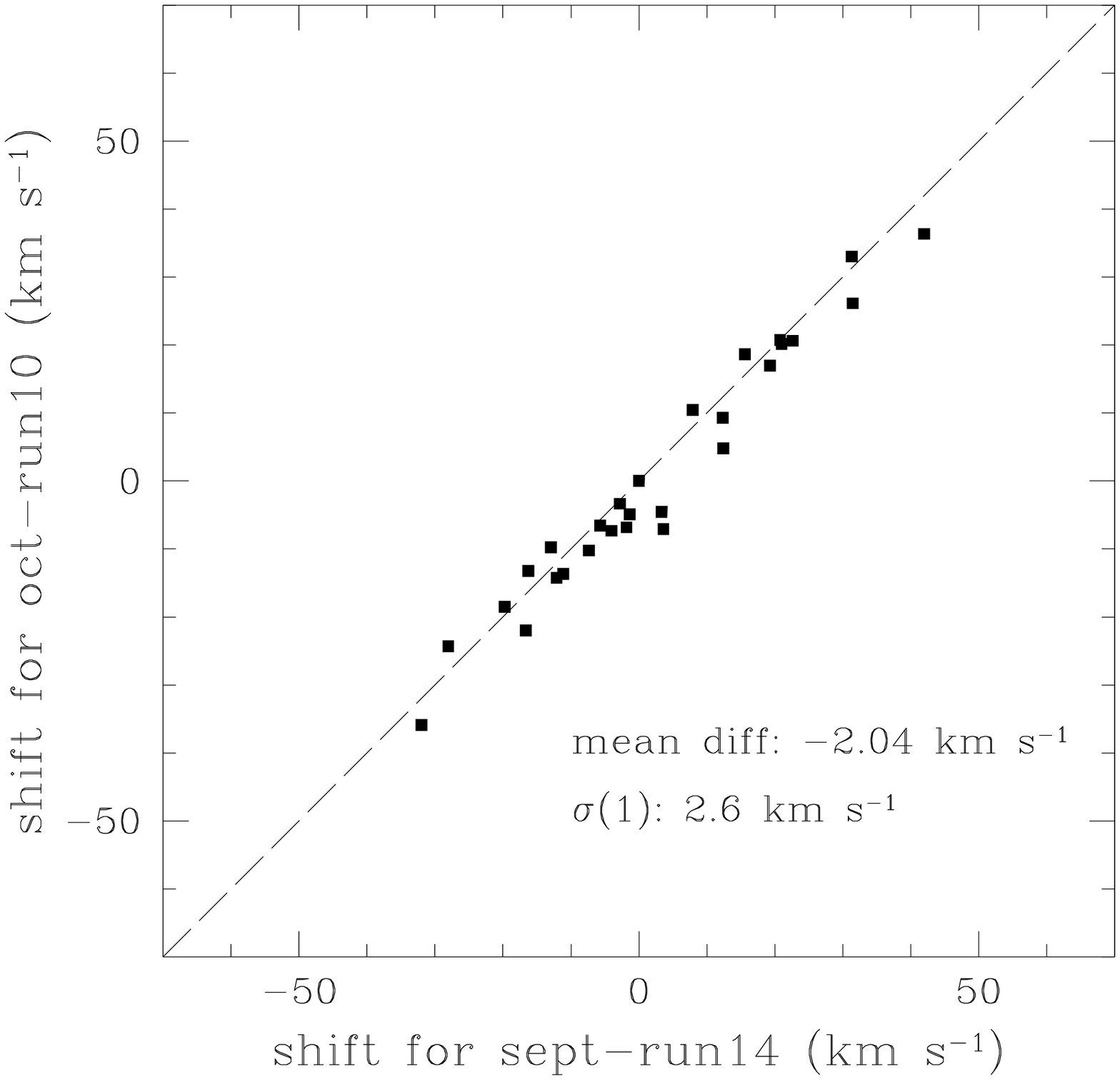}
\caption{Test observation of the radial velocity of stars performed in bright time 
by K. Freeman on 6dF (R=4000) in the second half of 2001. The figure correlates the
radial velocity measured with the same fiber (a) at two zenith distances (left), 
and (b) at two different epochs one month apart (right).}
\end{figure}

\begin{itemize}
\item {\bf The pilot survey} is a preliminary spectral survey, using
the existing 6dF system at the UK Schmidt telescope of the
AAO to observe about $100\;000$ stars in
$\approx 180$ days of unscheduled bright-time during the years
2003--2005. The 6dF spectrograph will target on the Ca triplet region 
 (850-875nm) favored by the GAIA instrument definition team. Spectra 
are taken at a resolution of $R=4000$. 
Test observations indicate that an accuracy of 2\,km\,s$^{-1}$ 
can be achieved (see Fig.~1).

The target list would include a large fraction of the $118\;000$ HIPPARCOS
stars that are accessible from the southern hemisphere as well as some
of the $2\;539\;913$ stars of the Tycho-2 catalog. The survey will focus
on stars in the color range $0.4 < B-V < 0.8$.  For these stars, useful
photometric parallaxes can be derived if the trigonometric parallaxes
are not available. The 6dF spectra will also provide useful estimates
of the $[\mbox{Fe/H}]$ value.
\item {\bf The main survey} will utilize a new Echidna-style multi
fiber spectrograph (see Fig.~2) at the UK-Schmidt telescope. It
consists of a 2250-spine fiber array covering the full field of the
Schmidt Telescope (40 deg$^2$). The spines are hexagonally-packed on
$\approx 7\,$mm centers. Each spine can be deformed by a piezo element
resulting in a 15-arcmin patrol area. The key advantage of this new
MOS design is its short reconfiguration time of $\approx
5$\,min. Echidna will feed an efficient spectrograph using a
high-efficiency Volume Phase Holographic (VPH) grating and will use a
single 2k by 4k red-optimized detector.  With 250\AA\ coverage
required for the Ca triplet region, a dispersion of 0.375\AA\
pixel$^{-1}$ will yield $R\approx 10\;000$ spectroscopy with 2-pixel
sampling.

Again the survey will target the Ca triplet region. At $R=10000$ we
expect radial velocities at 1\,km\,s$^{-1}$ accuracy. Iron abundances
$[\mbox{Fe}/H]$ could be determined to 0.1 dex accuracy and useful estimates
of the differential $[\alpha/\mbox{Fe}]$ abundance ratios for about half the
stars ($V<15$).

The main study is expected to be performed throughout the period
2006-2010. Owing to the short reconfiguration time of the new Echidna
MOS, about $22\;000$ stars can be observed per night in 30 min
exposures, resulting in S/N=30 spectra at $V=15$. Throughout the 5
year campaign, a total of 25 million stars can thus be
targeted. Subject to an equivalent instrument on the northern
hemisphere, RAVE will yield a total of 50 million radial velocities,
metallicites and, for a subset of $\la 50\%$ of the sample, also abundance ratios.

\end{itemize}

The RAVE survey will provide a vast stellar kinematic database
three orders of magnitude larger than any other survey proposed for
this decade. The main data product will be a magnitude-limited survey
of 26 million thin disk main sequence stars, 9 million thick disk
stars, 2 million bulge stars, 1 million halo stars, and a further 12
million giant stars including some out to 60 kpc from the Sun.  RAVE
will offer the first truly representative inventory of stellar radial
velocities for all major components of the Galaxy. 

\begin{figure}
\plotone{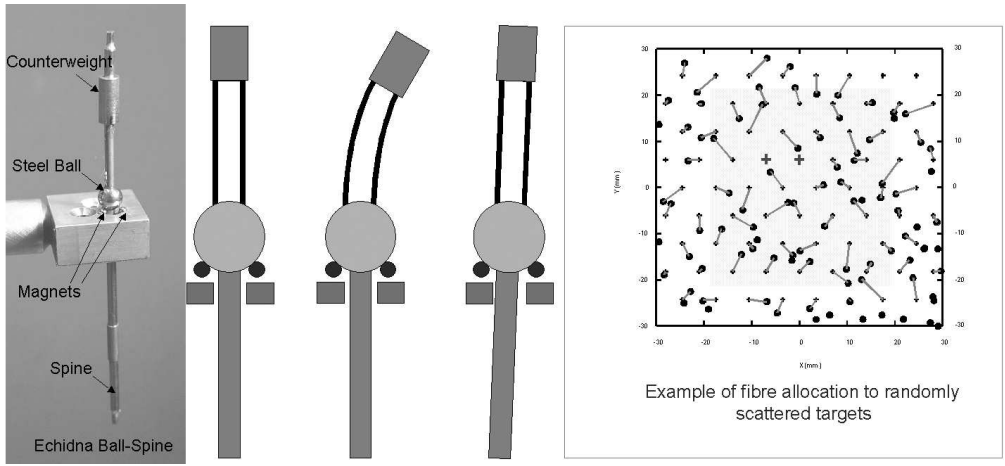}
\caption{The Echidna concept; left:  single ball-spine of the Echidna MOS; center:
illustration of the work principle of an individual echidna spine; 
right: example of fiber allocation to randomly scattered targets.}
\end{figure}

\section{The RAVE science case}

Within the cold dark matter (CDM) paradigm, the Galaxy built up
through a process of accretion over billions of years from the outer
halo. Sophisticated computer simulations of structure growth within a
CDM universe have now begun to shed light on how this process may have
taken place (see e.g. Steinmetz \& Navarro 2002). 
These advanced computer models do not only provide
information about the structure and kinematics of the major stellar
components of a galaxy but also on their chemical signatures and their
stellar age distribution. In the context of these simulations, RAVE 
will revolutionize our understanding of the
formation and evolution of all major components of the Galaxy: the
disk, the bulge and the halo.

\paragraph{Halo sub-structure.}  The details of galaxy formation are not well
understood. In particular, CDM simulations actually predict far more
infalling satellites than are currently observed 
(Moore et al. 1999; Klypin et al., 1999). The orbital
timescales of stars in the outer parts of galaxies are several billion
years and it is here we would expect to find surviving remnants of
accretion. The disrupting Sagittarius dwarf spheroidal galaxy
was discovered by Ibata et al.~(1994) from a multi-fiber radial
velocity survey. Five years later, Helmi et al.~(1999) discovered a stellar stream
within 1 kpc of the Sun after combining a radial velocity catalog
with the HIPPARCOS database.

Both of these studies demonstrate the power of accurate radial
velocities and proper motions in identifying cold stellar streams. We
can expect RAVE to reveal evidence of many tens of similar streams
both in the halo, in the outer bulge and within the thick disk. When
combined with the next generation of surveys such as DIVA, this tally
may extend to many hundreds of infalling systems.

\paragraph{Chemical signatures.}  A key aspect of RAVE will be the availability of 
chemical signatures like $[\alpha/\mbox{Fe}]$ and $[\mbox{Fe/H}]$, in addition to
accurate kinematics. The $\alpha$ elements arise from massive stars and the
bulk of their mass is released in Type II supernova explosions. The
Fe-peak elements are produced primarily by Type Ia supernovae which
begin to dominate after a billion years or more. Unique signatures
from abundance ratio pairs like ($[\alpha/\mbox{Fe}]$, $[\mbox{Fe/H}]$) may help to identify
a common site of formation among widely separated stars. The use of
chemical signatures can be extended to other components of the Galaxy,
in particular, the halo and the outer bulge and the thick disk.

\paragraph{Bulges.} The formation of stellar bulges, a major element of 
galaxy classification schemes, is not well understood. The Galactic
bulge stars are almost as metal rich as the thin disk but as old as
the thick disk and parts of the halo. Our current picture is that
large bulges are formed from a rapid collapse of a spherical cloud,
and that the small bulges are either formed from accretion or from the
action of the central bar after the disk formed. In an alternative
model, as favored by the CDM model of structure formation, bulges are
the remnants of early gas-rich mergers between some of the first
building blocks of a galaxy.  A key constraint is the $[\alpha/\mbox{Fe}]$
abundance ratio, which has been determined for only a few dozen bulge
stars to date. A short star formation epoch either during the
collapse of the bulge or, as favored in the CDM model, in the
progenitors from which a bulge is assembled is expected to lead to
enhanced $[\alpha/\mbox{Fe}]$ for most of the stars; an extended star
formation period during the bulge assembly would imply $[\alpha/\mbox{Fe}] =0$.

\paragraph{Thick disk.} The stars in the thick disk are at least as
old as those in the globular cluster 47 Tuc and it 
is widely believed to be a `snap frozen' relic of the early disk
shortly after the onset of disk dissipation. In this picture, an
infalling satellite vertically heated the early disk to a scale height
of 1 kpc. Another possibility is that the thick disk is made up of
tidal debris from infalling satellites.

From the combined chemical and kinematic signatures, the RAVE survey
should cleanly distinguish between competing models for the thick disk
and thus end a many-year old debate on the origin of the thick disk. A
major unknown in disk formation is whether the extent of the stellar
disk is laid down during the major epoch of dissipation, or whether it
grows with time. The RAVE survey will clearly establish whether the
radial extent of the thick disk is comparable to or less than the thin
disk. The chemical information will also be very important. The
existence of an abundance gradient in the thick disk, as we observe in
the thin disk, would argue against an infall origin; unique chemical
signatures in the thick disk would argue for an infall origin.

\paragraph{Thin disk.} The largest fraction of the RAVE targets will 
be thin disk dwarfs and giants. Little is known about the dynamical
state of the thin disk beyond 2 kpc of the Sun. The existence of the
inner bar and outer stellar warp is firmly established but many areas
of astrophysics would benefit from their influence being understood in
far more detail. Some external galaxies have optical disks, which
appear to be lop-sided with respect to the dark halo. Whether this is
the case for the Galaxy is not known. The intrinsic brightness of the
giants allows these important tracers to be observed throughout the
entire optical extent of the Galaxy. The giants probe the large-scale
dynamical state of the Galaxy, in particular the influence of the
inner bar, the outer warp and the degree of eccentricity and
lop-sidedness of the optical disk.

The vast number of dwarf stars in the RAVE survey will reveal the
dynamical state of the thin disk and neighboring spiral arms within a
few kiloparsecs of the Sun's position. This is crucial information if
we are to construct an accurate model of the gravitational potential
of the disk, and its distribution function. Recently, it has become
clear that even the old stellar populations appear to show
sub-structure. The RAVE survey will provide fundamental information on
how different stellar populations deviate from dynamical equilibrium,
and therefore constrain the formation history of the disk and its
different components (e.g. spiral arms, stellar associations, etc).

\subsection{Summary and Conclusions}
RAVE is an international project designed to survey radial velocities,
metallicities and abundance ratios for the brightest 50 million stars
in the Galaxy down to a completeness limit of V=16. This is
sufficiently deep to allow for kinematic and chemical studies of all
major stellar components of the Galaxy. RAVE has a number of science
goals addressing a wide range of priority areas in galactic structure
and dynamics. 

The RAVE survey is expected to be performed throughout the period
2003-2010. It also provides an opportunity to pre-empt some of the
spectral work in the GAIA mission, providing results up to one decade
earlier than that planned for the GAIA final release and probably still
well ahead of the launch of GAIA (2012). Thus RAVE serves as an ideal
real-data training set for the final design of the GAIA data reduction
pipeline and may even influence some of the final design decisions.

\end{document}